\newcommand{\delslash}{\partial \hspace{-6pt}/}
\newcommand{\la}{\langle}
\newcommand{\ra}{\rangle}
\newcommand{\lla}{\la\!\la}
\newcommand{\rra}{\ra\!\ra}
\newcommand{\beq}{\begin{eqnarray}}
\newcommand{\eeq}{\end{eqnarray}}
\renewcommand{\d}{\partial}

\renewcommand{\theequation}{\thesection.\arabic{equation}}

\newcommand{\chis}{ChS\,\,}
\newcommand{\pipi}{$\pi$-$\pi$\, }

\newcommand\con{\langle \bar {q} q \rangle }

\newcommand{\btem}{\bibitem}

\newcommand{\TK}{T.\ Kunihiro}
\newcommand{\MPL}{Phys.\ Lett.\ {\bf B}}
\newcommand{\MPTP}{Prog.\ Theor.\ Phys.}
\newcommand{\MPR}{Phys.\ Rev.}

\newcommand{\THK}{T. Hatsuda and T. Kunihiro}
\newcommand{\KH}{T. Kunihiro and T. Hatsuda}
\documentclass[a4paper,12pt,epsf]{article}
\usepackage{graphicx}
\begin{document}



\begin{center}
  Chiral Restoration and the Scalar and Vector 
Correlations in Hot and Dense Matter
\\

Teiji \textsc{Kunihiro}%
\\
Yukawa Institute for Theoretical Physics, Kyoto University,\\
Sakyoku, Kyoto 606-8502, Japan
\end{center}



\abstract{
First, it is pointed out that  hadron/nuclear physics
 based on QCD should  be regarded as ``condensed matter physics'' 
 of the QCD vacuum.
We indicate that phase shift analyses which respect 
chiral symmetry (\chis\hspace{-.1cm}), 
analyticity and crossing symmetry of the scattering 
amplitude show the $\sigma$ meson 
pole in the $s$-channel in the low mass region
as well as the $\rho$ meson pole in the $t$-channel 
in the \pipi scattering in the scalar channel.
We review recent developments
in exploring possible precursory phenomena of
partial  restoration of  \chis in nuclear medium by examining
 the spectral function in the scalar and the vector
channels. 
We emphasize that the wave function renormalization of the 
pion in the medium plays an essential role to induce the
decrease of the pion decay constant as the order parameter
of chiral transition. An emphasis is also put on the importance 
to examine the  scalar and vector
 channels simultaneously for exploring the possible restoration of chiral 
 symmetry.
}


\setcounter{equation}{0}
\renewcommand{\theequation}{\arabic{section}.\arabic{equation}}

\section{Introduction}

The basic observation on which the  whole discussions in this 
report are based is that the dynamical breaking of 
chiral symmetry (\chis\hspace{-.1cm}) is 
a phase transition of the QCD vacuum with an order 
parameter $\con $.
This is a reflection of the complicated structure of
QCD vacuum, which  actually makes
  hadron/nuclear physics based on  QCD tricky: 
\\
(1)\, The QCD Lagrangian 
 is not written in terms of  hadron fields but 
 in terms of  quark- and gluon-fields from which
 hadrons are composed, and 
the quarks and gluons are colored objects which 
can not exist in the asymptotic states.
 The low-lying elementary excitations
 on top of the non-perturbative QCD vacuum are composite
and  colorless particles, which we call hadrons.\\
(2)\, Symmetries possessed by the QCD Lagrangian,
such as the chiral SU(3)$_L\times$SU(3)$_R$ symmetry
 in the massless limit of the first three quarks, and 
the color gauge symmetry, 
are not manifest in our every-day world.
This complication is owing to the fact that 
the true QCD vacuum is completely different from the 
perturbative one and is actually realized through the phase 
transitions, i.e.,
 the confinement-deconfinement and the chiral transitions.
The notion of such a complicated vacuum structure, 
i.e., 
 the {\em collective} nature of the vacuum and the elementary particles
was first introduced by Nambu\cite{nambu}, in analogy with the
physics of superconductivity\cite{nambu2}.
\\
(3)\, Some phenomenological rules extracted on the hadron 
dynamics such as the vector-meson
dominance 
and the Okubo-Zweig-Iizuka 
rule might be 
 related with some fundamental properties of the QCD vacuum.
One may  notice that the so called $U_A(1)$ anomaly
which is responsible to make $\eta'$ as heavy as $960$ MeV
 also characterizes the non-perturbative QCD vacuum\cite{hk94}.\\ 
(4)\, 
Thus  one recognizes that hadron/nuclear physics as sub-atomic physics
based on QCD should be a study of the nature of QCD vacuum; i.e.,
hadron/nuclear physics  
 is a combination of the condensed matter physics of the 
QCD vacuum\cite{hk94}
 and the atomic physics as played with the constituent
quark-gluon model where the vacuum structure is taken for 
granted\cite{consti}.

In the present report, focusing on the
chiral transition in hot and/or dense hadronic matter,
I will discuss some characteristic changes in the scalar and vector
 correlations associated with
the (partial) restoration of \chis
 in the hadronic medium;
the major part of this report is based on 
some  previous ones\cite{protvino,yitp00,paris}.

\setcounter{equation}{0}
\renewcommand{\theequation}{\arabic{section}.\arabic{equation}}

\section{Restoration of Chiral Symmetry as  a Phase 
Transition of QCD vacuum}

The lattice simulations\cite{karsch}  show that
the true QCD vacuum is realized through 
the chiral transition.
At finite $T$, the lattice simulations also show that \chis 
is restored at $T=T_c\sim $150-175 MeV, depending the number of the
active light flavors.
A heuristic argument based on 
a Hellman-Feynman theorem can tell us that the chiral condensate 
$\con$ decreases at finite density $\rho_{_B}$ as well as at 
finite $T$.

The chiral condensate at $T \neq 0$ is given by
\beq
 \lla \bar{q}_iq_i \rra   = \frac{1}{Z}
 {\rm Tr} \left[ \bar{q}_iq_i\   e^{- (H_{QCD}-\mu N)  /T} \right] 
  =  \frac{\partial \omega(T)}{\partial m_{i}}
\equiv \la \bar{q_i}q_i\ra_0
+\delta \lla \bar{q_i}q_i\rra_T,
\eeq
where $Z$ is the QCD partition function,
 $\omega$ is the  free energy density and $m_i$ the current quark 
mass ($i=u, d, s, \dots$).
For the free pion gas, the modification due to finite $T\not=0$
is given by
\beq
\delta \lla \bar{q_i}q_i\rra_T
=\sum_pn_{\pi}(p)\la\pi(p)\vert\bar{q}_iq_i\vert\pi(p)\ra,
\eeq
 with
$\la\pi(p)\vert\bar{q}_iq_i\vert\pi(p)\ra
=m_{\pi}\la \bar{q}_iq_i\ra_{\pi}/E^{\pi}_p,$
being the pion matrix element of the scalar charge $\bar{q}_iq_i$
due to the i quark. The non-covariant matrix element
$\la \bar{q}_iq_i\ra_{\pi}$ is given by the Hellman-Feynman 
theorem, again;
$\la \bar{q}_iq_i\ra_{\pi}=\d m_{\pi}/\d m_i$.
At small temperature, it can be shown that the 
pion matrix element of the scalar charge 
$\la\bar{q_i}q_i\ra_{\pi}\simeq 6.25>0$ with the use of
 Gell-Mann-Oakes-Renner relation,
$f_{\pi}^2m_{\pi}^2=-(m_u+m_d)/2\cdot\la \bar{u}u+\bar{d}d\ra$.
Thus, one may expect that
$\lla\bar{q_i}q_i\rra$
decreases in the absolute value;
 restoration of \chis at finite temperature.
More systematic calculations using a nonlinear chiral Lagrangian confirm
 the above result\cite{gerber}.

As for the the degenerate nucleon system $\vert N\ra$,
one may start from the formula\cite{DL90},
\beq
 \langle N\vert \bar{q}q\vert N\rangle=
\frac{\partial\langle N\vert {\cal H}_{QCD}\vert N\rangle}
{\partial m_q},
\eeq
where the expectation value of QCD Hamiltonian may be evaluated to 
be\\ 
$ \langle N\vert {\cal H}_{QCD}\vert N\rangle=
\varepsilon_{vac}+\rho_{_B}[M_N+B(\rho_{_B})]$.
Here, $\varepsilon_{vac}, M_N$ and $B(\rho_{_B})$ denote the vacuum 
energy, the nucleon mass and the nuclear binding energy per particle,
respectively.
Thus one ends up with
\beq
{\langle \bar{q} q \rangle \over
 \langle \bar{q} q \rangle_{0} }
= 1 - {\rho_{_B} \over f_{\pi}^2 m_{\pi}^2 } \left( \Sigma_{\pi N}
 + \hat{m} {d \over d\hat{m}}  B(\rho_{_B}) 
\right) ,
\eeq
where
$\Sigma _{\pi N}=(m_u+m_d)/2\cdot\la N\vert \bar{u}u+\bar {d} d\vert 
N\ra$ denotes the  $\pi$-N sigma term with $\hat{m} =(m_u + m_d)/2$;
the semi-empirical value of $\Sigma _{\pi N}$ is known to 
be $(40 - 50)$ MeV.
Notice that the correction term with finite $\rho_{_B}$ is positive
and  gives a reduction of almost 35 \% of $\langle \bar{q} q \rangle $
already at the normal nuclear matter density $\rho_0 = 0.17 $fm$^{-3}$. 
We notice that the physical origin of this reduction is common
in finite-$T$ and -$\rho_{_B}$ cases: The scalar probe $\bar{q}_iq_i$
 hits either the vacuum or a particle h present in the system at
$T\not=0$ and/or $\rho_{_B}\not=0$;
in the latter case, $\bar{q}_iq_i$ picks up a positive contribution
to the chiral condensate  
because of the positive scalar charge $\la h | \bar{q}q | h \ra >0$
of the particle.

From the above estimate, one may consider that
 the central region of heavy nuclei could be dense enough to  cause 
a partial restoration of \chis, realizing
some characteristic phenomena of the chiral restoration in nuclear
medium; 
some of them may be observed by  experiments 
in the laboratories on Earth\cite{hk94,BR}.

It is a well-known fact in many-body or statistical physics that
if a phase transition is of second order or weak first order,
there may exist specific  collective excitations called 
 {\em soft modes}\cite{soft}; they actually correspond to 
the quantum fluctuations of the order parameter.
In the case of chiral transition, 
there are two kinds of fluctuations; 
those of the phase and the modulus of the chiral condensate. 
The former is the Nambu-Goldstone boson, i.e., the pion, while 
the latter the $\sigma$  with the quantum numbers $I=0$ and $J^{PC}=0^{++}$.
Some effective models\cite{ptp85}
 of QCD and the argument based on the 
``mended symmetry'' of Weinberg\cite{mended} predict the $\sigma$ mass 
below or equal to the $\rho$ meson mass.

\setcounter{equation}{0}
\renewcommand{\theequation}{\arabic{section}.\arabic{equation}}

\section{Low-energy QCD and the $\sigma$ meson}

\subsection{The $\sigma$ and
 chiral symmetry, unitarity, analyticity and crossing symmetry
 in the $\pi$$\pi$ scattering amplitudes}

The elusiveness of the $\sigma$ meson comes from the fact that 
it strongly couples 
to two pions to acquire  a large width 
$\Gamma \sim m_{\sigma}$.
After the establishment of the chiral perturbation theory
\cite{chipert} for
describing low energy hadron phenomena,
one of the most important problems has been to describe resonances
in a consistent way with \chis\cite{OOR}.
The central issue there is to incorporate 
the fundamental properties of the scattering amplitude
such as   unitarity,
 analyticity and the crossing symmetry together with \chis. 
In this way, the  recent cautious phase shift
 analyses of the \pipi scattering have come to 
claim a  pole identified with
 the $\sigma$ in the $s$ channel together with the $\rho$ meson
pole in the $t$ channel\cite{pipiyitp,pipi,CLOSETORN}.
The $\sigma$ pole has  the real part 
Re\, $m_{\sigma}= 500$-600 MeV and the imaginary 
part Im\, $m_{\sigma}\simeq {\rm Re}\, m_{\sigma}$\cite{CLOSETORN}.
Afterwards, it has been  also found that
 the $\sigma$ pole gives a significant contribution in 
 the decay processes of heavy particles involving a charm and 
$\tau$ leptons;
as D$\to \pi \pi \pi$,
$J/\psi \to \sigma\omega \to 2\pi\omega$ and $\tau\to a_1\nu \to \sigma\pi\nu \to 3\pi\nu$ 
 \cite{E791,bes,cleo}.

A  summary of the locations of the $\sigma$ pole in 
the complex energy plane may be found in \cite{XZ01}.

To see the significance in establishing the 
$\sigma$ pole of incorporating \chis,
analyticity and crossing symmetry as well as unitarity,
we notice  that 
the same phase shift can be well 
reproduced with a unitarized scattering amplitude 
lacking the $\sigma$ pole but including the
$\rho$ meson pole in the $t$-channel\cite{juerich,isgur}.
One may naturally wonder 
whether the existence of  the $\sigma$ pole in the $s$ channel is
 real or not.
It was apparent that the urgent problem is  to 
incorporate  crossing symmetry in the scattering amplitudes consistently
with \chis.
Then Igi and Hikasa\cite{igi} constructed the invariant amplitude for the 
\pipi scattering using the $N/D$ method 
so that it satisfies the \chis low energy 
theorem, analyticity, unitarity and especially 
(approximate) {\em crossing symmetry}.
They calculated  two cases with and without the scalar 
pole degenerated with the $\rho$ meson, the existence of which
was taken for granted.
What they found is that the $\rho$ only scenario can account only 
about a half of the observed phase shift, while the degenerate
$\rho$-$\sigma$ scenario gives a reasonable agreement with the
data. In the phenomenological approaches like those given in 
\cite{juerich}, it was  unclear how the fundamental properties 
of the scattering matrix are respected, such as
 \chis and  crossing symmetry.

\subsection{The $\sigma$ as a collective $q$-$\bar{q}$ mode on top of
the QCD vacuum}

In fact, the interpretation of the $\sigma$ has a long history of 
controversies\cite{CLOSETORN}:
First of all, the conventional constituent quark model
 has difficulties to describe 
the low-lying $\sigma$ as a $q$-$\bar{q}$ meson; the lowest
 scalar $q$-$\bar{q}$ meson should be a $P$ wave ($^3P_0$) state, 
which is 
in turn usually heavier than 1 GeV.
Then there are various proposals for interpretation
of  the low-mass $\sigma$, 
commonly denying a pre-existing 
$q$-$\bar{q}$ state.
However, it should be emphasized that the $\sigma$ as the quantum
fluctuation of the chiral order parameter
must be a {\em collective} state composed of many $q$-$\bar{q}$ states
 as the pion is\cite{nambu,ptp85}; notice that the pion can not be understood
within the conventional constituent quark model.

In \cite{yitp00},
it is argued that 
the linear realization of \chis as given in the 
NJL-like dynamical models, especially incorporating
 the vector terms 
is consistent with the chiral perturbation theory in the sense
 that they reproduce the low-energy constants $L_i$ and $H_i$
in the nonlinear chiral Lagrangian\cite{bijnens}.

\subsection{Possible roles of the $\sigma$ in hadron phenomenology} 

If the $\sigma$ meson with a low mass has been identified, many 
 experimental facts  which otherwise are mysterious 
 might  be nicely accounted 
for  in a simple way\cite{hk94,ptp95}.
The phenomena which have a possbile relevance to the $\sigma$ include:
(1) ~ $\Delta I=1/2$ rule in the kaon decay\cite{morozumi},
(2)~ the intermediate-range attraction in nuclear force
\cite{sawada},
 (3) ~  $\pi$-N sigma term\cite{JK,kuni90} and so on.
They are related to the collective nature of the $\sigma$, which
causes an enhancement of the matrix elements involving the $\sigma$
in the intermediate state.

\setcounter{equation}{0}
\renewcommand{\theequation}{\arabic{section}.\arabic{equation}}
\section{Partial chiral restoration and the 
$\sigma$ meson in  hadronic matter}

As remarked above, it still remains uncertain whether
the $\sigma$ pole reported 
really corresponds to the quantum fluctuation of the 
chiral order parameter or only a $\pi$-$\pi$ molecule generated
dynamically without the pre-existing $\sigma$.
As was  first shown in\cite{HK85}, 
if the $\sigma$ is really associated
with the fluctuation of the chiral order parameter,
one can  expect that the $\sigma$ pole moves toward
the origin in the complex energy plane
in the chiral limit and the $\sigma$ may become a sharp
resonance as \chis is restored 
 at high temperature and/or density; the $\sigma$ can be
a {\em soft mode}\cite{soft} of the chiral restoration; see also
\cite{bernard87}.

One must, however, notice that 
 a hadron put in a heavy nucleus may 
 dissociate into complicated
excitations to loose its identity.
Then the most proper quantity to observe is the response function 
or spectral function in the channel with the same quantum number 
as the hadron has. 
When
 the coupling of the hadron  with the environment is relatively small,
then there may remain a peak with a small width in the spectral 
function, 
corresponding to the hadron, thereby it may be meaningful to tell
the mass and the width of the hadron in the medium.

How about at finite density?
The lattice QCD is unfortunately 
still premature to give any reliable results for the system 
at finite chemical potential.
If numerical experiment may not be relied on,
 one can listen to Nature directly.
Some years ago\cite{tit,ptp95}, the present author
proposed several nuclear experiments including
 one using electro-magnetic probes 
to create the scalar mode in nuclei, thereby 
obtain a clearer evidence of  the  existence of 
the $\sigma$ meson and also examine the possible 
restoration of chiral  symmetry in  nuclear
 medium. It was also mentioned that
to avoid the huge amount of two pions from the $\rho$ meson, 
detecting  neutral pions through four $\gamma$'s may be convenient.

Hatsuda, Shimizu and the present author (HKS) \cite{HKS}
showed that the spectral enhancement
near the $2m_{\pi}$ threshold takes place 
in association with  partial restoration of \chis  
at finite baryon density.
The calculation is a simple extension of the 
finite $T$ case done by Chiku and Hatsuda\cite{CH98}:
  Chiku and Hatsuda performed a model calculation of
 the spectral functions in
the pion and the $\sigma$ meson channel at finite $T$ using a
linear $\sigma$ model.
They found that the spectral function in the 
$\sigma$ channel shows an enhancement near two-$m_{\pi}$ threshold as 
\chis is restored as $T$ goes high.

In \cite{HKS}, HKS started from  the following linear sigma model;
\beq
\label{model-l}
{\cal L}&=&   {1 \over 4} {\rm Tr} [\partial M \partial M^{\dagger}
 - \mu^2 M M^{\dagger} - {2 \lambda \over 4! } (M M^{\dagger})^2
 -  h (M+M^{\dagger}) ]
  + \bar{\psi} ( i \delslash - g M_5 ) \psi\nonumber \\
 & &  + \cdots  ,
\eeq
where  
$M = \sigma + i \vec{\tau}\cdot \vec{\pi}$,
$M_5 = \sigma + i \gamma_5 \vec{\tau}\cdot \vec{\pi}$,
 $\psi$ is the nucleon field, and
 Tr denotes the trace for the flavor index.

The spectral function in the scalar channel is
obtained from the propagator. 
The $\sigma$-meson propagator at rest in the medium reads
$D^{-1}_{\sigma} (\omega)= \omega^2 - m_{\sigma}^2 - 
\Sigma_{\sigma}(\omega; \rho_{_B})$,
where $m_{\sigma}$ is the mass of the $\sigma$ in the tree-level, and
$\Sigma_{\sigma}(\omega; \rho_{_B})$ represents
the loop corrections
in the vacuum as well as in the medium.
 The corresponding spectral function is given by 
$\rho_{\sigma}(\omega) = - \frac{1}{\pi} {\rm Im} D_{\sigma}(\omega)$.

Now one can easily verify that 
${\rm Im} \Sigma_{\sigma}
\propto \theta(\omega - 2 m_{\pi})
 \sqrt{1 - {4m_{\pi}^2 \over \omega^2}}$
near the two-pion threshold  in the one-loop order.
On the other hand,
the pole mass $m_{\sigma}^*$ in the medium 
is  defined by
${\rm Re}D_{\sigma}^{-1}(\omega = m_{\sigma}^*)=0$.
Partial restoration of \chis 
implies that $m_{\sigma}^*$
  approaches to $ m_{\pi}$.  
Thus one sees that  there should exist a density $\rho_c$ at which 
 ${\rm Re} D_{\sigma}^{-1}(\omega = 2m_{\pi})$
 vanishes even before the complete restoration
 of \chis where $\sigma$-$\pi$
 degeneracy gets realized;
${\rm Re} D_{\sigma}^{-1} (\omega = 2 m_{\pi}) =
 [\omega^2 - m_{ \sigma}^2 -
 {\rm Re} \Sigma_{\sigma} ]_{\omega = 2 m_{\pi}} = 0$.
At this point, the spectral function is solely 
given in terms of the
 imaginary part of the self-energy;
\beq
\rho_{\sigma} (\omega \simeq  2 m_{\pi}) 
 =  - {1 \over \pi \ {\rm Im}\Sigma_{\sigma} }
 \propto {\theta(\omega - 2 m_{\pi}) 
 \over \sqrt{1-{4m_{\pi}^2 \over \omega^2}}},
\eeq
which clearly shows the near-threshold enhancement of
the spectral function.
 This should be a general phenomenon to be realized
in association with  partial restoration of \chis.

\begin{figure}[t]
\centering
\includegraphics[scale=.38]{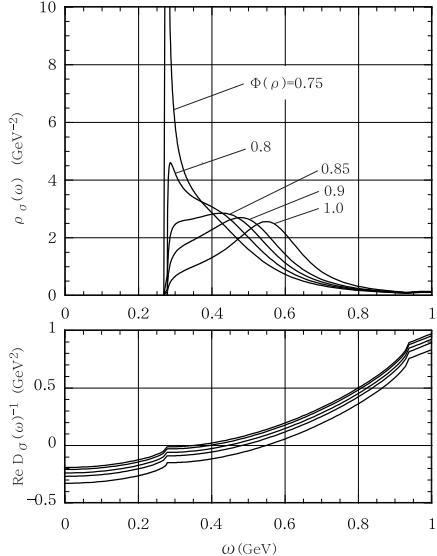}
\label{fig.1}
\caption{The spectral function $\rho_\sigma(\omega)$ (the upper panel)
and ${\rm Re} D_{\sigma}^{-1}(\omega)$  (the lower panel)
 calculated with a linear sigma model.
$\Phi(\rho)\equiv \langle \sigma \rangle/ \sigma_0$ measures
 the rate of the partial restoration of the \chis at the
baryonic density $\rho$.\cite{HKS}
}
\end{figure}

In fact, 
 the effects of the meson-loop as well as
 the baryon density  was treated as a perturbation 
 to the vacuum quantities in the above. 
Therefore, our loop-expansion  
  should be valid only at relatively low
 densities.
When we parameterize the chiral condensate in nuclear matter
 $\langle \sigma \rangle$ as
$\langle \sigma \rangle \equiv  \sigma_0 \ \Phi(\rho)$,
 one may take the linear density approximation for 
small density;
$\Phi(\rho) = 1 - C \rho / \rho_0 $,
with
$C = (g_{\rm s} /\sigma_0 m_{\sigma}^2) \rho_0$.

The spectral function $\rho_\sigma(\omega)$ together with 
${\rm Re} D_{\sigma}^{-1}(\omega)$  
 calculated with a linear sigma model are shown 
  in Fig.1 with the bare mass $m_\sigma=550$ MeV:
 The characteristic enhancements of the spectral
 function is seen just above the 2$m_{\pi}$.
It is also to be noted that even well before
 $m_{\sigma}^*$ is close to $m_{\pi}$, 
when \chis is restored,
 a large enhancement
 of the spectral function is seen near $2m_{\pi}$ threshold.

\setcounter{equation}{0}
\renewcommand{\theequation}{\arabic{section}.\arabic{equation}}
\section{Chiral restoration in nonlinear realization;
 role of the wave function renormalization}

The near-threshold enhancement obtained above
is based on the linear representation of the
 \chis, where the $\sigma$ degree of freedom is 
explicit from the outset.
Actually, some related works use the non-linear realization of
\chis supported by the development of the chiral 
perturbation theory;
a unitarized  chiral perturbation theory is intended to
be applied to nuclear medium, which is a far from  simple task.
Nevertheless it would be intriguing to see how the possible 
restoration of \chis is implemented in the 
nonlinear realization of \chis where the
$\sigma$ degrees of freedom is absent.

Jido et al\cite{jhk}
showed that the nonlinear realization of the chiral 
symmetry can also
give rise to a near 2$m_{\pi}$ enhancement of the
spectral function in nuclear medium as shown in Fig.2.
The enhancement of the cross section is found due to 
the wave function renormalization of the pion in nuclear medium
which causes the decrease of the pion decay constant $f_{\pi}^{\ast}$
in the medium.

\begin{figure}[tb]
\centering
\includegraphics[scale=.4]{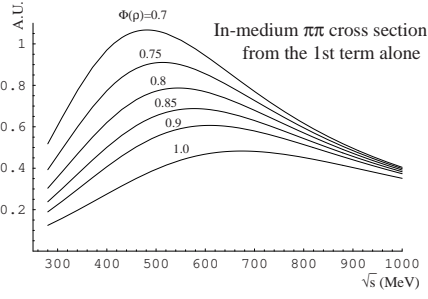}
\caption{
In-medium $\pi\pi$ cross section in the $I=J=0$ channel
 in the heavy $S$ limit where $m_{\sigma}^*$ is taken to be infinity.
    The cross section is shown in the arbitrary unit 
(A.U.).\cite{jhk}
 }
\label{fig2}
\end{figure}

Jido et al also start with the Lagrangian (\ref{model-l}), and 
take a polar decomposition of the chiral field,
$M = \sigma + i \vec{\tau} \cdot \vec{\pi} = (\la \sigma \ra + S) U$,
 with 
$U = \exp (i \vec{\tau} \cdot \vec{\phi} /f^{*}_{\pi})$.
Here $f^{*}_{\pi}$ is the would-be ``in-medium pion decay constant''.
Then (\ref{model-l}) is cast into the following form
\beq
\label{model-nl}
{\cal L}& =&   {1 \over 2} [(\partial S)^2 - m_{\sigma}^{*2} S^2]
  - {\lambda \la \sigma \ra \over 6} S^3 - {\lambda \over 4!} S^4 
  \nonumber \\
&  &
+  {(\la \sigma \ra +S)^2 \over 4} {\rm Tr}
 [\partial U \partial U^{\dagger}]  + { \la \sigma \ra + S \over 4}\ 
 h \ {\rm Tr}[U^{\dagger}+U] \nonumber \\
 && +
{\cal L}_{\pi N}^{(1)} - g S \bar{N} N \ .
\eeq
Here
${\cal L}_{\pi N}^{(1)} =
\bar{N}(i \delslash + i v \hspace{-6pt}/
 + i  a \hspace{-6pt}/   \gamma_{5}  - m_{N}^* ) N$,
and
$^{t}(v_{\mu},a_{\mu}) = (\xi \partial_{\mu} \xi^{\dagger}
 \pm \xi^{\dagger} \partial_\mu \xi)/2$,
with $m_N^* = g \la \sigma \ra$.
In this representation, the in-medium $\pi\pi$ scattering
amplitude reads in the tree level
\beq
\label{scatt2}
A(s) =  {s- m^2_\pi \over \la \sigma \ra^2}
       - {(s - m_\pi^2)^2 \over
 \la \sigma \ra^2} {1 \over s - m_{\sigma}^{*2}}  .
\eeq
The first term in (\ref{scatt2}) comes from the
 contact 4$\pi$ coupling generated by  the
 expansion of the second line in (\ref{model-nl}) with
 the coefficient proportional to
 $1 / {\la \sigma \ra}^2$.
 On the other hand,
 the second term in (\ref{scatt2}) is from the contribution of the
  scalar meson $S$ in the $s$-channel.
Fig.2 shows a unitarized in-medium $\pi\pi$ cross section solely
 with the first term in (\ref{scatt2}), i.e.,  the 
non-linear chiral Lagrangian. One can see a clear enhancement
  of the cross section near the threshold or a softening,
associated with restoration of  \chis.

Although there is no explicit $\sigma$-degrees of freedom in this 
heavy $S$ approximation, 
 there arises a decrease of the
 pion decay constant $f_{\pi}^{\ast}$ in nuclear medium.
Precisely speaking, the heavy $S$ limit is defined as
$\lambda$ and hence $m^{\ast}_{\sigma}$ go infinity with
$g/\lambda$ and $\la \sigma \ra_0=f_{\pi}$ fixed. 
Then integrating out the $S$ field in this limit,
one has the following low-energy Lagrangian
\beq
\label{model-nl2}
{\cal L} & = &
 \left(
 {f_{\pi}^2 \over 4}
 - {g f_{\pi}  \over 2 m_{\sigma}^2}\bar{N}N
 \right)
\left(
{\rm Tr} [\partial U \partial U^{\dagger}]
 - {h \over f_{\pi}} \  {\rm Tr}[U^{\dagger}+U] \right)
 \nonumber \\
 &  & \ + \  {\cal L}_{\pi N}^{(1)} + \cdot \cdot \cdot \ \ ,
\eeq
where all the constants take their vacuum values:
 $f_{\pi} = \la \sigma \ra_0 $,
 $m_{\sigma}^2 = \lambda \la \sigma \ra_0^2/3 + m_{\pi}^2$,
 and  $m_N = g \la \sigma \ra_0$.
 Note that $g f_{\pi} / 2  m_{\sigma}^2$
 in front of $\bar{N}N$ approaches to
 a finite value
 $3 g/2\lambda f_{\pi}$ in the heavy limit, thus it cannot be neglected.
 In (\ref{model-nl2}),  $\cdot \cdot \cdot$
  denotes
  other higher dimensional operators which are not relevant for the
   discussion below.

It should be remarked that the
 near-threshold enhancement is caused by the decrease of 
$f^{\ast}_{\pi}$ 
 owing to the following  new vertex:
\beq
\label{new-vertex}
{\cal L}_{\rm new} = - {3g \over 2 \lambda f_{\pi}} \
\bar{N}N {\rm Tr} [\partial U \partial U^{\dagger}].
\eeq
 In Fig.3,  4$\pi$-$N$-$N$ vertex generated by ${\cal L}_{\rm new}$
 is shown as an example.
 In the uniform nuclear matter, $\bar{N}N$ in
 eq.(\ref{new-vertex}) may be replaced by $\rho$;
$ \bar{N}N\, \rightarrow \rho$.

 Then 
the coefficient of the first term of 
eq.(\ref{model-nl2}) is written 
as 
$f_{\pi}^2/ 4\cdot(1 - 2g/f_{\pi}  m_{\sigma}^2\cdot\bar{N}N)$,
which shows that 
 the properly  normalized  pion field  in nuclear matter must 
be
\beq
 \phi^{\ast} = (\phi /f_{\pi}) \cdot f_{\pi}^* .
\eeq
with
$ f_{\pi}^*=f_{\pi} (1- g  \rho/f_{\pi} m_{\sigma}^2)$.
 The in-medium $\pi\pi$ scattering
 with this normalization exactly reproduces
  the first term in (\ref{scatt2}) as it should be.
It is important that the wave function renormalization 
of the pion field in the medium
implies  
 a reduction of  the vacuum condensate
\beq
f_{\pi} = \la \sigma \ra_0\,  \rightarrow f_{\pi}^*= \la \sigma \ra .
\eeq
The essential role of the  wave-function renormalization of the pion field
 on  the partial restoration of \chis 
 is  confirmed  rigourously in the chiral perturbation theory
\cite{mow} and has been also 
  revealed  in accounting for the anomalous repulsion seen in the
deeply bound pionic nuclei\cite{kkw}. 

 Here it should be pointed out that
 the vertex eq.(\ref{new-vertex})
 which  we have extracted from the linear sigma model
 has been known to be one of the next-to-leading order terms
 in the non-linear chiral Lagrangian in the
  heavy-baryon formalism  \cite{GSS}.
Although roles of the new vertex  have  been extensively discussed 
in \cite{mow} in the context of chiral perturbation theory,
 the full non-perturbative incorporation of it, however,
has not been done so far  especially in  calculations
of  the $\pi\pi$ scattering amplitudes in nuclear matter starting
from  the non-linear chiral Lagrangian\cite{wambach}.


\begin{figure}[tb]
\centering
\includegraphics[scale=.5]{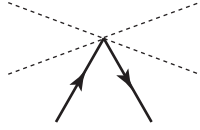}
\caption{
The new 4$\pi$-$N$-$N$ vertex generated
 in the nonlinear realization.
 The solid line with arrow and the dashed line represent
 the nucleon and pion, respectively.\cite{jhk}}
\label{fig.3}
\end{figure}

\section{Concluding remarks}
Interestingly enough,
CHAOS collaboration  \cite{chaos}
 observed that
the   yield for  $M^A_{\pi^{+}\pi^{-}}$ 
 near the 2$m_{\pi}$ threshold 
 increases dramatically with increasing $A$;
 this experiment was motivated to explore the $\pi$-$\pi$
correlations in nuclear medium\cite{motivation}. They
identified that the $\pi^{+}\pi^{-}$ pairs in this range of
 $M^A_{\pi^{+}\pi^{-}}$ is in the $I=J=0$ state.
A similar experimental result which shows a softening of the spectral
function in the $\sigma$ channel has been also obtained by 
TAPS group\cite{taps}.

It is interesting that there are other possible 
experimental evidences for partial chiral restoration in 
nuclear matter than the chiral fluctuations in the
sigma meson channel discussed so far.
  The deeply bound pionic atom has proved 
to be a good probe of the properties of the hadronic interaction
 deep inside of heavy nuclei.  It has been 
 suggested \cite{WEISE,itahashi,kkw} that
the anomalous energy shift of the pionic atoms (pionic nuclei)
owing to the strong interaction could be attributed to the
 decrease of the effective pion decay constant 
$f^{\ast}_{\pi}(\rho)$ at finite density $\rho$, which 
may imply that the \chis is partially restored deep
inside of nuclei.
It is interesting enough that the decrease of $f^{\ast}_{\pi}(\rho)$
is owing to  the wave-function
renormalization of the pion field in nuclei\cite{kkw},
as mentioned before.
A KEK experiment also shows the softening of the spectral
function in the vector ($\rho/\omega$) channel in heavy nuclei such as
 Cu \cite{ozawa}, which might indicate also 
that restoration of \chis in nuclear medium as was suggested
in \cite{vector}.

Recently, Yokokawa et al\cite{yokokawa} have applied the $N/D$ method 
a la Igi and Hikasa to hot and dense matter and 
 examined the pole structures of the scattering 
matrices and  the spectral functions in the $\sigma $ 
and
the $\rho$ meson channels on the equal footing:
 The effect of chiral restoration
is taken into account in the mean field level, which 
is  tantamount
to replacing $f_{\pi}$ by $f_{\pi}^*$.
It was found that there exist two kinds of poles 
in the complex energy plane in both
channels in the scattering matrix, one of which moves toward the
origin in the chiral limit; this implies especially that 
 the sigma meson which is elusive in the free space may
appear as a rather sharp resonance in hot and/or dense medium
 where \chis is partially restored. This pole behavior was
first suggested in \cite{jhk} and shown in \cite{paris}.
Yokokawa et al have also shown that the spectral functions in
the both channels  give rise to a softening
 in tandem as the \chis is restored.

More than a decade ago, the present author \cite{kuni91}
examined characteristic
behaviors of the baryon 
number susceptibility $\chi_B=\d \rho/\d \mu$
 at finite temperature and density.\footnote{
In lack of time, I failed to present this part in the workshop.}
It was shown that the $\chi_B$ is actually the correlation function 
in the vector channel (the longitudinal component) and coupled to 
the scalar susceptibility $\d \la \bar{q}q\ra/\d m$ at finite
chemical potential $\mu\not=0$, and hence the latter can reflect the
singular behavior of the former around the critical point of the 
chiral transition.
It certainly be fruitful to examine the scalar and vector
 correlations in a combined way to explore possible critical
phenomena of the chiral transition in hot and/or dense medium.

\section*{Acknowledgements}
The major part of this report is  based on the work
done in 
collaboration with T. Hatsuda,
  H. Shimizu, D. Jido and K. Yokokawa.
I thank them for the collaboration and  discussions. 
This work is supported by the Grants-in-Aids of the Japanese
Ministry of Education, Science and Culture (No. 14540263). 

%

\end{document}